\documentclass[aps,prb,twocolumn,preprintnumbers,amsmath,amssymb,superscriptaddress]{revtex4-1}

\usepackage{times}
\usepackage{graphicx}% Include figure files
\usepackage{dcolumn}% Align table columns on decimal point
\usepackage{bm}% bold math%\documentclass[aps,prl,twocolumn,groupedaddress]{revtex4-1}
\usepackage{natbib}
\usepackage{hyperref}
\hypersetup{backref,pdfpagemode=UseOutlines,colorlinks=true}%
\makeindex

\begin{document}
%\preprint{}
\title{Germanium-based quantum emitters towards \textit{time-reordering entanglement scheme}  with degenerate exciton and biexciton states}

%%%%%%%%%%%%%%%%%%%%%%%%%%%%%%%%%%%%%%%%%%%%%%%%%%%%%%%%%%%%%%%%%%%%%
\author{Nicola Dotti}
\affiliation{LENS, Dipartimento di Fisica, Universit\`a di Firenze, Via Sansone 1, I-50019 Sesto Fiorentino, Italy}
%%%%%%%%%%%%%%%%%%%%%%%%%%%%%%%%%%%%%%%
%%%%%%%%%%%%%%%%%%%%%%%%%%%%%%%%%%%%%%%
\author{Francesco Sarti}
\affiliation{LENS, Dipartimento di Fisica, Universit\`a di Firenze, Via Sansone 1, I-50019 Sesto Fiorentino, Italy}
%%%%%%%%%%%%%%%%%%%%%%%%%%%%%%%%%%%%%%%
%%%%%%%%%%%%%%%%%%%%%%%%%%%%%%%%%%%%%%%
\author{Sergio Bietti}
\affiliation{L-NESS and Dipartimento di Scienza dei Materiali, Universit\`a di Milano Bicocca, Via Cozzi 53, I-20125 Milano, Italy}
%%%%%%%%%%%%%%%%%%%%%%%%%%%%%%%%%%%%%%%
%%%%%%%%%%%%%%%%%%%%%%%%%%%%%%%%%%%%%%%
\author{Alexander Azarov}
\affiliation{Department of Physics, University of Oslo
NO-0316 Oslo (Norway)}
%%%%%%%%%%%%%%%%%%%%%%%%%%%%%%%%%%%
%%%%%%%%%%%%%%%%%%%%%%%%%%%%%%%%%%%%
\author{Andrej Kuznetsov}
\affiliation{Department of Physics, University of Oslo
NO-0316 Oslo (Norway)}

%%%%%%%%%%%%%%%%%%%%%%%%%%%%%%%%%%%%%%
%%%%%%%%%%%%%%%%%%%%%%%%%%%%%%%%%%%%%%%
\author{Francesco Biccari}
\affiliation{LENS, Dipartimento di Fisica, Universit\`a di Firenze, Via Sansone 1, I-50019 Sesto Fiorentino, Italy}

%%%%%%%%%%%%%%%%%%%%%%%%%%%%%%%%%%%%%%
%%%%%%%%%%%%%%%%%%%%%%%%%%%%%%%%%%%%%%%
\author{Anna Vinattieri}
\affiliation{LENS, Dipartimento di Fisica, Universit\`a di Firenze, Via Sansone 1, I-50019 Sesto Fiorentino, Italy}
%%%%%%%%%%%%%%%%%%%%%%%%%%%%%%%%%%%%%%%
%%%%%%%%%%%%%%%%%%%%%%%%%%%%%%%%%%%%%%%
\author{Stefano Sanguinetti}
\affiliation{L-NESS and Dipartimento di Scienza dei Materiali, Universit\`a di Milano Bicocca, Via Cozzi 53, I-20125 Milano, Italy}
%%%%%%%%%%%%%%%%%%%%%%%%%%%%%%%%%%%%%%%
%%%%%%%%%%%%%%%%%%%%%%%%%%%%%%%%%%%%%%%

\author{Marco Abbarchi}
\thanks{Corresponding author: Marco Abbarchi (marco.abbarchi@im2np.fr)}  %\email{}
\affiliation{CNRS, Aix-Marseille Universit\'e, Centrale Marseille, IM2NP, UMR 7334, Campus de St. J\'er\^ome, 13397 Marseille, France}

%%%%%%%%%%%%%%%%%%%%%%%%%%%%%%%%%%%%
%%%%%%%%%%%%%%%%%%%%%%%%%%%%%%%%%%%%%

\author{Massimo Gurioli}
\thanks{Corresponding author:  Massimo Gurioli (gurioli@fi.infn.it)} %\email{}
\affiliation{LENS, Dipartimento di Fisica, Universit\`a di Firenze, Via Sansone 1, I-50019 Sesto Fiorentino, Italy}

\date{\today}

\begin{abstract}

We address the radiative emission of individual germanium extrinsic
centers in  Al$_{0.3}$Ga$_{0.7}$As epilayers grown on  germanium
substrates.  Micro-photoluminescence
experiments demonstrate the capability of high temperature emission (70
K) and complex exciton configurations (neutral exciton X and biexciton XX, 
positive X$^{+}$ and negative X$^{-}$ charged exciton) of these quantum emitters.
Finally, we investigate  the renormalization of each  energy level showing a
large and systematic change of the binding energy of XX and X$^{+}$ from
positive to negative values (from $\sim$+5 meV up to $\sim$-7 meV covering
about $\sim$ 70 meV of the emission energy) with increasing quantum
confinement. These light emitters, grown  on a silicon substrate,  may exhibit 
energy-degenerate X and XX energy levels. Furthermore, they  emit at the highest 
detection efficiency window of Si-based single photon detectors. These features 
render them a promising device platform for the generation of entangled photons in 
the \textit{time-reordering scheme}.

\end{abstract}
\pacs{}
%\keywords{}
%\maketitle
%
%
\maketitle

\section{INTRODUCTION}\label{sec:INTRO}

The implementation  of quantum states of light is at the base of most
quantum computation and quantum information
protocols:\cite{Imamog1999,Bennett2008,Nielsen2010} bright and high-quality
single-, entangled- and indistinguishable-photon emitters are a necessary
resource for quantum key distribution, quantum repeaters and photonic
quantum information
processing.\cite{Li2003,Fattal2004,Politi2009,Troiani2000} In the same
field, other important applications of solid-state  quantum-sources consist
in the possibility to couple them with atomic vapours for the production of
quantum-bit  based on slow-light memories\cite{Akopian2011,Siyushev2014} or
on electron and holes spin.\cite{De2012,Gao2012,Warburton2013}
Single photon emission\cite{Lounis2005,Eisaman2011} has been demonstrated
in several solid state systems such as epitaxial and colloidal quantum
dots,\cite{Chen2000,Zwiller2001,Regelman2001a,Stievater2001,Brokmann2004,
Stevenson2006,Kumano2006,Abbarchi2009a,Kuroda2009,Ulhaq2012,Trotta2012a,
Kuroda2013,Jayakumar2013,Trotta2014,Birindelli2014}  carbon nanotubes\cite{Hogele2008} and
single molecules.\cite{Basche1992,Brunel1999,Lounis2000} 

%Moreover,
%exploiting a resonant laser excitation scheme, the coherence of a single
%photon emitter can be extended at will,  enabling the distribution over
%large distances of single and indistinguishable photons through conventional
%optical fibers.\cite{Nguyen2011,Jayakumar2013,Proux2014}
 
More advanced implementations of quantum states of light, such as entangled
photon pairs, can be obtained in solid state systems, provided the binding
of two correlated electronic states  within the same
nanostructure.\cite{Chen2000,Li2003,Stevenson2006,Kuroda2013,Jayakumar2013,
Trotta2012,Trotta2014}  Several schemes for the generation of entangled
photons have been proposed for quantum dots. Most of them are based on the
neutral biexciton-exciton cascade XX-X: provided the implementation  of a
spin-degenerate neutral exciton transition,  a maximally entangled photon
pair  can be encoded in the polarization degree of XX and X photons. The
requirement of spin degeneracy can be satisfied for negligible electron-hole
spin interaction allowing to erase the \textit{which-path information} in
the biexciton-exciton cascade. This kind of spin-degenerate state  has been
implemented  either as an  \textit{a priori}, built-in characteristic of the
nanostructure (like in Reference \onlinecite{Kuroda2013} where highly
symmetric and unstrained quantum dots were
grown\cite{Mano2010,Ha2014,Liu2014}), by spectral
filtering the photons having the same energy\cite{Akopian2006} or by tuning
\textit{a posteriori} the fine interaction to zero.
\cite{Stevenson2006,Trotta2012a,Trotta2012,Trotta2014}

A different protocol called \textit{time-reordering scheme}, has been
recently proposed in order to implement polarization entanglement in the
emitted photon cascade (XX-X) from QDs with arbitrary
spin-splitting.\cite{Troiani2008,Avron2008}  This scheme is based on the
zero biexciton binding energy allowing to erase the  \textit{which-path
information} by introducing \textit{a posteriori} and \textit{ad-hoc} delay
of the XX-X emitted photons. The time-reordering protocol relaxes the need
of a perfectly spin-degenerate neutral exciton state, but the condition of
zero biexciton binding energy is not straightforward to be realized, and
this photon entanglement scheme has not been yet experimentally
demonstrated. Still there are a few reports on QDs  naturally exhibiting
degenerate XX and X states\cite{Rodt2003,Rodt2005,Rodt2007,Schliwa2009} or
on the possibility of tuning \textit{a posteriori} the XX binding energy to
zero with an external
control.\cite{Trotta2012a,Trotta2012,Trotta2014}
 
%Finally, we mention that, provided a bright and sharp single photon source
%of indistinguishable photons,\cite{Nguyen2011,Proux2014} different
%algorithms can be implemented for obtaining entangled photons
%pairs\cite{Jayakumar2013} such as the
%\textit{time-bin-entanglement},\cite{Simon2005,Halder2007,Larque2008,Martin2
%013,Sheng2013}  or with an entangling CNOT gate obtained by exploiting
%linear-optical components.\cite{Knill2001,Gazzano2013} These evidences are
%of the utmost importance as they completely  relax the constraints on the
%electronic properties of the light source while shifting the
%\textit{entangling mechanism }  to an external device (such as a Franson
%interferometer or a CNOT gate) [IO TOGLIEREI, PORTA FUORI DAL TARGET].
% 
%%Finally, in very recent reports, the use of   sharp transitions from
%charged excitons in a quantum dot has been proposed as a valuable way of
%measuring temperature of an electronic reservoir as low as 100
%mK.\cite{Seilmeier2014,Haupt2014} [QUEST'ULTIMA FRASE SI PUO' ANCHE
%TOGLIERE....]
 
A second relevant route to obtain quantum light sources in semiconductor
devices in alternative to conventional QDs, is related to the exploitation
of extrinsic centers in III-V, IV and II-VI semiconductor
alloys.\cite{Thomas1966,Castell2003,Klingshirn2007,Kenraad2011} Single
photon emission from isolated impurity centers has been shown  in
ZnSe/ZnMgSe alloys,\cite{Sanaka2009} tellurium isoelectronic dyads in
ZnSe,\cite{Muller2006}  nitrogen impurity centers in
GaAs\cite{Francoeur2004,Ikezawa2012,Zhang2013,Ethier2014} and  in
AlAs\cite{Jo2013} and with  nitrogen-vacancy centers
and chromium in
diamonds.\cite{Kurtsiefer2000,Beveratos2002,Jacques2007,Castelletto2010,
Ikezawa2012,Lesik2013,Spinicelli2011}
 
Within this class of quantum emitters, some impurity centers (for example dyads 
complexes\cite{Muller2006,Jo2013}) allow to
confine exciton complexes, thus leading to the possibility to be exploited 
as sources of entangled photons. Nevertheless,
for extrinsic centers,  the scientific literature on XX-X cascade is much
less flourished with respect to QDs and, for example, the possibility to obtain
time-reordering has not yet been reported. Among several interesting
features, a peculiar properties of extrinsic centers is the possibility to
exploit them in indirect band-gap based devices, such as carbon\cite{Cloutier2005,Berhanuddin2012} and copper\cite{Weber1982,Sumikura2014} impurity centers in
silicon  and carbon antisite-vacancy pairs in
SiC.\cite{Castelletto2014} This last example has been demonstrated to be a
bright single photon emitter at room temperature.  

In a recent paper, we showed a hybrid III-V/IV-IV single photon device based on extrinsic emitters in
Al$_{0.3}$Ga$_{0.7}$As\cite{Minari2012} fabricated with a low thermal budget
method on Ge and Si substrates, thus providing a solid-state platform towards 
the integration of quantum light sources in \textit{classical} electronic
devices.

In this  paper we firstly unambiguously demonstrate, by a comparative analysis with samples grown in different conditions, the connection of these latter extrinsic emitters\cite{Minari2012} with Ge contamination of the
Al$_{0.3}$Ga$_{0.7}$As alloy. Then, by a careful micro-photoluminescence analysis we are able to imaging a large sample region 
isolating a large ensemble of extrinsic defect and determining their density and \textit{pseudo-macro-photoluminescence} spectrum. We also perform a
statistical analysis of the excitonic complexes from these Ge-centers, demonstrating 
the change from binding to
anti-binding of the biexciton XX and positive charged exciton X$^{+}$
states with the increase of the emission energy of the corresponding neutral
exciton transition X, thus opening the possibility to implement a
\textit{time reordering scheme} for XX-X photon cascade exploiting defects in
semiconductors. We finally validate their photostability at high temperature and
show a selective quantum-confined-Stark effect as the main origin of the
different inhomogeneous line broadening of the s-shell bright states  (the
two neutral states  X and  XX, plus the positive and negative charged
excitons, X$^{+}$ and X$^{-}$ respectively).  It is important noting that 
the devices in use were grown on standard Ge substrates and they emit in the 
spectral interval of highest detection efficiency of most single-photon silicon-based 
light-detectors.

The paper is organised as follows: in Section \ref{sec:EXPERIMENTAL} we
provide a description of the experimental setup used for the
micro-photoluminescence (micro-PL) investigation and the sample fabrication; In Section
\ref{sec:$Al$-segregation vs $Ge$-centers} we precisely address the origin
of the extrinsic centers as Ge related defects, comparing purely III-V
(Ge-free samples) with  samples grown on a germanium substrate. In  addition 
we map the Ge-centers
emission over large areas thus enabling the
precise isolation of the related \textit{pseudo-macroPL}
ensemble emission. In Section
\ref{sec:HIGH POWER}   the high-power excitation regime is discussed showing
the onset of multiexcitonic and charged exciton features. We also 
discuss the photostability at high temperature of individual Ge-centers
as well as the quantum confined Stark-effect originated from charged defects
in the surrounding semiconductor matrix.   Finally, in Section
\ref{sec:BINDING ENERGY}, we discuss the features of the binding energy of
XX, X$^{+}$ and X$^{-}$ showing the transition from binding to anti-binding
of XX and X$^{+}$. In Section \ref{sec:CONCLUSIONS} we draw the conclusions.

\section{EXPERIMENTAL: SAMPLE PREPARATION AND EXPERIMENTAL
METHODS}\label{sec:EXPERIMENTAL}

\begin{figure}[t]
   \begin{center}\leavevmode
   \includegraphics[width=.8\linewidth]{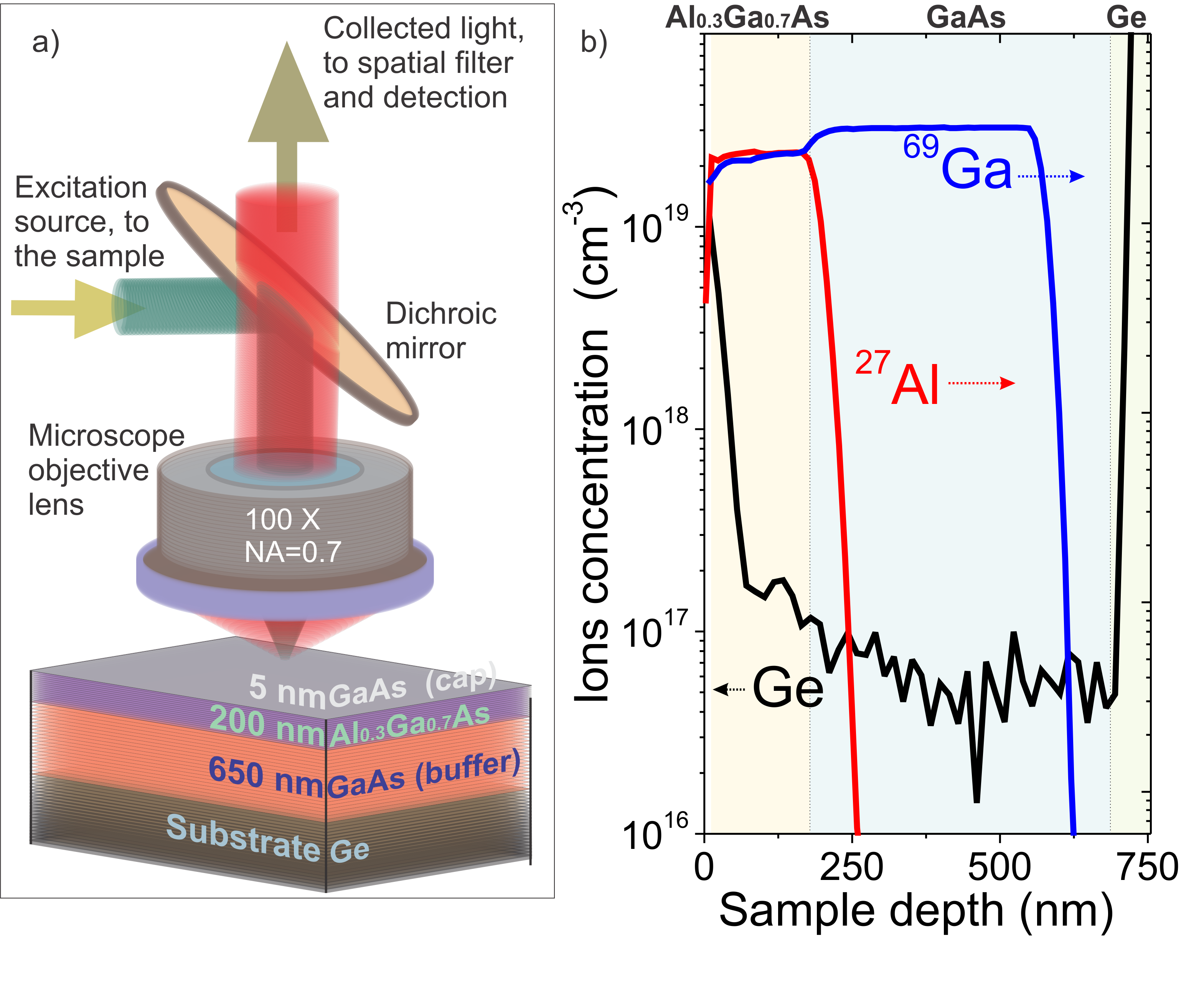}% Here is how to import EPS art
   \caption{ a) Bottom, sketch of the sample composition: 5 nm GaAs (capping layer), 200 nm Al$_{0.3}$Ga$_{0.7}$As  (active layer), 650 nm GaAs (buffer layer), Ge (substrate). Top, 
   scheme of the optical  apparatus used for PL experiments. b) SIMS measurements on the investigated sample. Ge (left axis, calibrated), Ga and Al (right axis, not calibrated) ions concentration are plotted in a logarithmic scale as a function of the milled depth. Shaded areas highlight the different layers. \label{fig:Fig1}}
   \end{center}
   \end{figure}

A thorough description of the device in use can be found in reference
\onlinecite{Minari2012,Sarti2013} where  first evidences of excitonic
recombination and single photon emission were shown for Ge-impurities in
Al$_{0.3}$Ga$_{0.7}$As grown on germanium and silicon substrates. The
results found for the two different kind of substrates are very similar and
here we concentrate our attention only on the Ge-substrate case. A scheme of the sample cross section is given in the
bottom part of  Fig.\ref{fig:Fig1}a) (this sample will be denoted as Ge 580 $^{\circ}$C in the following).
 
%For a proper growth of the III-V layer,   avoiding the formation of
%antiphase domain,  the germanium substrates were miscut by 6 degrees towards
%the [100] direction. A scheme of the sample cross section is given in the
%bottom part of  Fig.\ref{fig:Fig1} a): in a molecular beam epitaxy machine,
%an initial GaAs buffer 650 nm was grown on the germanium substrate at 580$^{\circ}$C
%with a Ga deposition rate of 0.76 ML/s. A 200 nm thick Al$_{0.3}$Ga$_{0.7}$As 
%layer was deposited on the GaAs (1.07 ML/s deposition rate) at the same
%substrate temperature used for the GaAs buffer deposition. Finally a 10 nm
%thick GaAs top layer caps the structure for avoiding the
%Al$_{0.3}$Ga$_{0.7}$As oxidation. A post-growth rapid thermal annealing (4 minutes at 800$^{\circ}$C in nitrogen atmosphere) was used for reducing the intrinsic defectivity typically present in samples grown below 580 $^{\circ}$C.\cite{Sanguinetti2008}

As references, we grew two test-samples with similar parameters of
those described before but on conventional GaAs substrates and, in one case,  
also using low temperature growth in order to facilitate possible Al clustering. 
%thus completely
%excluding the presence of Ge-contamination (this is done to test the
%possible emission from nanometer-sized, Al-poor clusters in the
%Al$_{0.3}$Ga$_{0.7}$ matrix which may lead to similar features in the
%emission spectrum,\cite{Heiss2013,Corfdir2014} see Section
%\ref{sec:$Al$-segregation vs $Ge$-centers} for a detailed discussion). 
For the first test sample on GaAs we grew a 200 nm Al$_{0.3}$Ga$_{0.7}$As layer at  580 $^{\circ}$C for the first 100 nm, 400 $^{\circ}$C for the central 30 nm and again  580 $^{\circ}$C for the last 70 nm (this sample will be denoted as GaAs 400 $^{\circ}$C in the following). In the second test sample the temperature was set at  580 $^{\circ}$C for the full 200 nm Al$_{0.3}$Ga$_{0.7}$As thickness (this sample will be denoted as GaAs 580 $^{\circ}$C in the following).

%Macro-photoluminescence experiments were performed in a cold finger cryostat at
%10 K. The samples were excited with a CW laser (532 nm solid state laser)
%on a circular spot as large as 100 $\mu$m in diameter. For
For micro-PL experiments the samples were kept at low temperature
in  a low-vibration liquid He-flow cold-finger cryostat which in turn  was
mounted on a stepping motor translation-stage for scanning the sample surface. A
schematic view of the experimental setup is shown in  the top part of
Fig.\ref{fig:Fig1} a). More details are given in reference
\onlinecite{Minari2012,Sarti2013}

%: a high resolution microscope objective (100$\times$
%magnification, numerical aperture 0.7) focused the excitation laser
%source  on a $\sim$5 $\mu$m spot on the sample surface. The same microscope
%objective was used to collect the photoluminescence emission which, after a
%dichroic mirror, was focused on a spatial filter. In this  case the spatial
%filter was a monomode optical  fiber with a core diameter of 5 $\mu$m
%leading to a lateral resolution of about 1 $\mu$m in diameter at the
%relevant working wavelength. At the exit of the fiber the collected light
%was then fed into a monochromator (a 30 cm single-grating Czerny-Turner or a
%100 cm double-grating Jobin Yvon monochromator) and detected by a Si-based
%CCD camera. The spectral resolution was about 180 $\mu$eV for the 30 cm
%spectrometer and 30 $\mu$eV for the 100 cm spectrometer case.
 
%With this experimental setup we  probed the emission of the sample by
%scanning the surface and collecting a spectrum for each step. We used this
%method for linear scans, thus building  one dimensional spectral maps
%(energy vs space), and two dimensional maps for hyper-spectral
%imaging.\cite{Bao2012} In order to exemplify this method  examples of spectrally resolved and animated 2D maps are shown  in these references %\onlinecite{VideoMap1,VideoMap2}. The details of each 1D 
%and 2D scans will be given through the text.

In the case of the Ge 580 $^{\circ}$C sample, the Ge concentration versus depth 
profiles was measured by secondary ion
mass spectrometry (SIMS) with a Cameca IMS 7f microanalyzer (the Ge
detection limit was estimated in $\sim$5 $\times$ 10$^{16}$ cm$^{-3}$). 
A 10 keV O$^{2+}$ primary beam with a current of 600 nA was rastered over a
150$\times$150 $\mu$m$^{2}$ area and secondary ions were collected from the
central part of the sputtered crater. The intensity-concentration
calibration was performed using Ge ion implanted samples as a reference. The
conversion from sputtering time to sample depth was performed by measurement
of the crater depth using a Dektak 8 stylus profilometer and assuming a
constant erosion rate.

The results of SIMS measurements are shown in Fig.\ref{fig:Fig1} b) in a
logarithmic scale as a function of the milled depth. A large Ge
concentration is found in both the GaAs buffer and Al$_{0.3}$Ga$_{0.7}$As layer, denoting a
Ge diffusion from the substrate into the MBE grown layers. 
%Note that the
%Ge-concentration increase at the surface is likely an artefact of the
%measure. 
Here the Ge contamination reaches the value
of $\rho_{Ge} \simeq  10^{17}$ cm$^{3}$ (not very different from the values
reported in references \onlinecite{Chand1986,Sieg1998} for similar samples),
sufficiently large  to form an impurity band. Indeed, according to the Mott
criterion for GaAs, the critical doping  $n_{c}$ to form a band   (defined
as $a_{B} n_{c} ^{1/3} = 0.25$, where $a_{B}$  is the Bohr radius),   is
$n_{c} \simeq 1.3 \times  10^{16} $cm$^{-3}$.\cite{Zhang2013} The relatively
large Ge contamination of  the Al$_{0.3}$Ga$_{0.7}$As alloy is an important
clue for the attribution of extrinsic quantum emitters to Ge-centers, which
will be discussed in the following. The large increase of the Ge content near 
the surface (first 50 nm, see Fig. \ref{fig:Fig1} b)) is likely due to 
segregation of Ge on the GaAs surface during the growth and/or
Ge out-diffusion towards the surface.

\section{Nature of the defects and ensemble spectra}\label{sec:$Al$-segregation vs $Ge$-centers}

  \begin{figure}[h]
  \begin{center}\leavevmode
  \includegraphics[width=.8\linewidth]{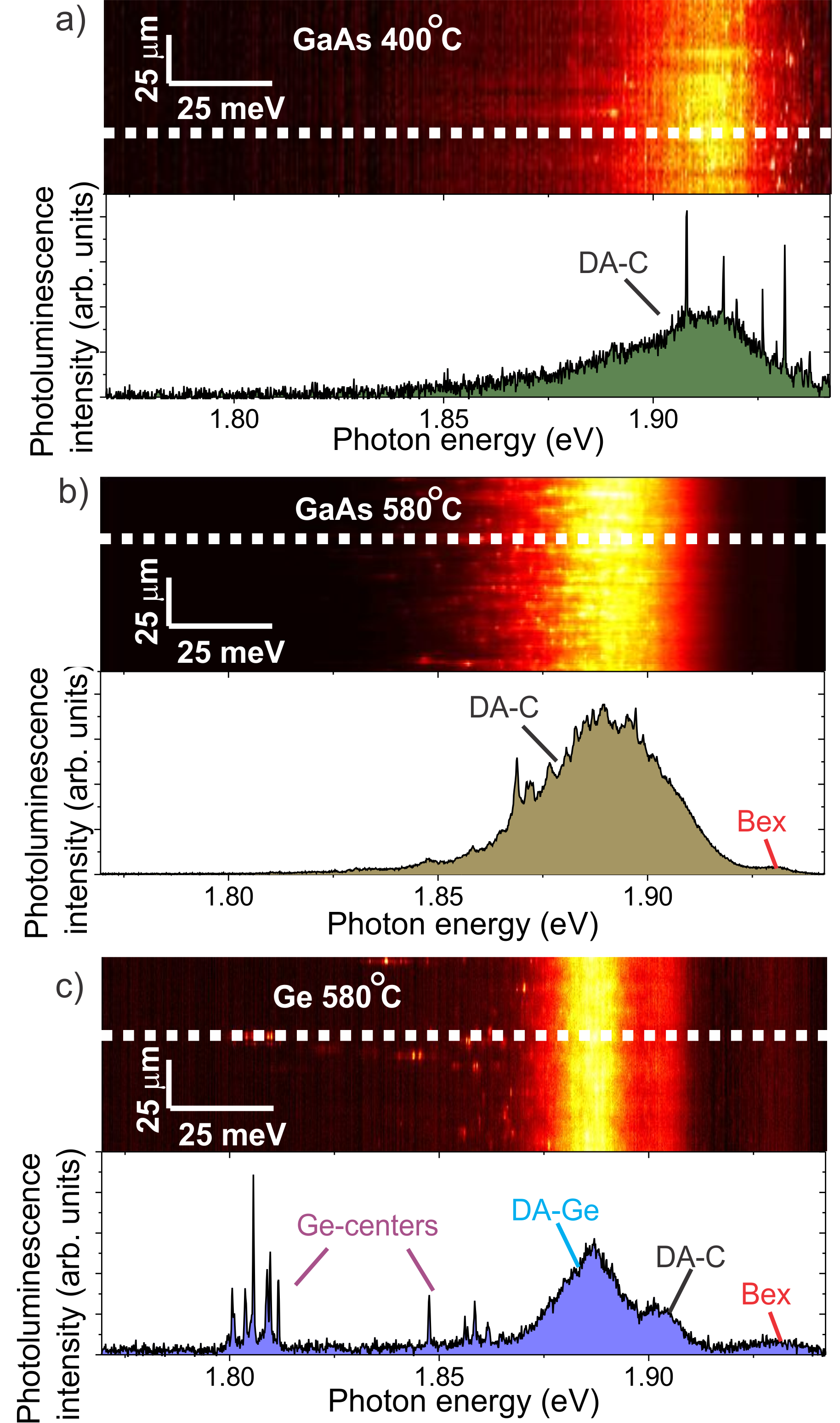}% Here is how to import EPS art
  \caption{ a), b) and c) Top panel shows a 1D scan over 100 $\mu$m (in 0.5$\mu$m steps) on samples GaAs 400$^{\circ}$C, GaAs 580$^{\circ}$C and Ge 580$^{\circ}$C. The white dashed line highlights the spatial position of the spectrum shown in the corresponding bottom panel. Bottom panels: typical PL spectrum extracted from the 
  1D scan in the top panel. \label{fig:Fig2}}
  \end{center}
  \end{figure}

Before addressing the optical properties of the extrinsic quantum emitters
in Al$_{0.3}$Ga$_{0.7}$As we unambiguously validate their origin as due to Ge
centers by comparing the emission of purely III-V samples with that of
those  grown on silicon and germanium. This is also done in order to exclude
the presence of Al-poor clusters in the  Al$_{x}$Ga$_{1-x}$As matrix. In
fact, the presence of such Al-poor zones 
has been recently proposed\cite{Heiss2013,Corfdir2014} as possible
explanation  of bright and sharp PL lines (emitting from
$\sim$1.7 eV up to $\sim$2 eV) in Al$_{x}$Ga$_{1-x}$As
nanowires.\cite{Minari2012,Sarti2013} Note that the excitonic
emission from those clusters\cite{Heiss2013,Corfdir2014}  falls in a similar
spectral interval of the Ge-defects emission investigated here (emitting
from $\sim$1.8 eV up to $\sim$1.87 eV).
 
In order to promote some alloy disorder, possibly producing  Al-poor
nanoclusters within the Al$_{0.3}$Ga$_{0.7}$As layer, we grew the two reference samples GaAs 400$^{\circ}$C  and
GaAs 580$^{\circ}$C. The micro-PL spectra
at low temperature (10 K) of the  two test samples are  compared with the
Al$_{0.3}$Ga$_{0.7}$As layer grown on Ge substrate in 
Fig.\ref{fig:Fig2}.  1D maps (energy vs space), with the detected
PL intensity represented as a false-color scale, are shown for each of 
the three samples. The top panel of Fig.\ref{fig:Fig2} a),b) and c) displays the full scan 
(100 $\mu$m long) while the bottom panel shows a typical spectrum collected in a single point.
In all the three samples we can identify  the bound
exciton emission (Bex) and the usual band from shallow centers due to
carbon-related residual contamination (DA-C).\cite{Veta1985,Pavesi1994}  This latter band is usually present
in AlGaAs layers grown by molecular beam epitaxy.\cite{Ploog1981,Veta1985,Oelgart1989,Pavesi1994} By comparing the
BEx and DA-C PL bands in the three spectra, we conclude that the Al content
is about 5$\%$ larger in the GaAs 400$^{\circ}$C sample with respect to the
580$^{\circ}$C sample. At the same time the Al content of the Ge 580$^{\circ}$C sample is
slightly larger (of the order of 1$\%$) than the GaAs 580$^{\circ}$C sample. Apart
from these unintentional differences in the calibration of the Al$_{x}$Ga$_{1-x}$As
alloy, we note a relevant structuring of the DA-C band with sharp lines
spatially localised, together with an overall reduced emission efficiency
for the lower growth temperature and a broader extension of the DA-C band
when compared with a similar sample grown at 650$^{\circ}$C (not shown).

Quite peculiar is the PL spectrum of the sample  grown on a Ge substrate
where, besides the presence of the same PL structures (DA-C
band and BEx  band), we observe   two other contributions at lower energy
(see Fig. \ref{fig:Fig2} c) and previously reported data in references
\onlinecite{Minari2012,Sarti2013}). These two new PL
contributions appear as a broad PL band at about $\sim$1.88 eV together
with, at lower energies, a tail extending till $\sim$1.80 eV  where several
sharp, isolated and bright lines organised in \textit{multiplets} are found.
Due to the relevant Ge contamination determined by the SIMS measurements reported in Fig.\ref{fig:Fig1} b),
the broad band at around $\sim$1.88 eV is ascribed to Ge-related
donor-acceptor recombination (it will be denoted as  DA-Ge band in the
following).\cite{Ploog1981,Veta1985,Oelgart1989,Pavesi1994}  Similarly to
the case of the DA-C band, also the DA-Ge band  may show sharp and spatially
localised lines. Finally and accordingly with the previous literature, the
tail extending to lower energy is interpreted as emission from deep
donor-acceptor levels.
\cite{Mooney1990,Pantelides1992,Feichtingerdeep,Klingshirn2007}
 
%We stress that, the sharp and localised features  in the DA-C and DA-Ge
%bands (observed in all the samples) correspond to individual and spatially
%localized lines (confirmed by power dependent measurements which does not
%lead to correlated additional contributions). On the contrary, when Ge is
%present, most of the bright emitting centers around and below 1.88 eV
%clearly show a \textit{multiplet structure} accounting for the confinement
%of multi-excitonic complexes (see also Section \ref{sec:HIGH POWER}).

In two recent papers, we showed that the sharp lines emissions below the
DA-Ge band show antibunching and biexciton recombination.\cite{Minari2012,Sarti2013} The comparison with Ge free samples and
the data from SIMS measurements, demonstrate that the Al-poor nanoclusters,
leading to natural QDs in the Al$_{x}$Ga$_{1-x}$As alloy (as observed in Ref
\onlinecite{Heiss2013,Corfdir2014}), do  not play a role in the attribution
of the sharp lines organised in localised multiplets which are observed only
in the sample grown on the Ge substrate. This new findings lead us to conclude
that the extrinsic centers under investigation have to be ascribed to the
presence of Ge contamination in the Al$_{0.3}$Ga$_{0.7}$As layer.

%These spectral intervals will
%be used in order to analyse the 2D spectral maps obtained in
%micro-photoluminescence experiments (see  Fig.\ref{fig:Fig3} and \ref{fig:Fig5} and the additional 
%animated maps obtained with a sharper spectral filter in references \onlinecite{VideoMap1,VideoMap2}). 
%It is worth noting that in macro-photoluminescence experiments the PL band associated to the Ge-defects
%emission in the region I is not well visible\cite{Minari2012}. 
%Therefore, for a precise analysis of these centers micro-photoluminescence 
%measurements are needed.

%\section{EMISSION MAPS AND ENSEMBLE SPECTRUM}\label{sec:LOW POWER}

%  I, II and III shaded areas highlight the
% spectral interval over  which we integrate the spectral maps shown in b).
 
\begin{figure}[t]
 \begin{center}\leavevmode
\includegraphics[width=.8\linewidth]{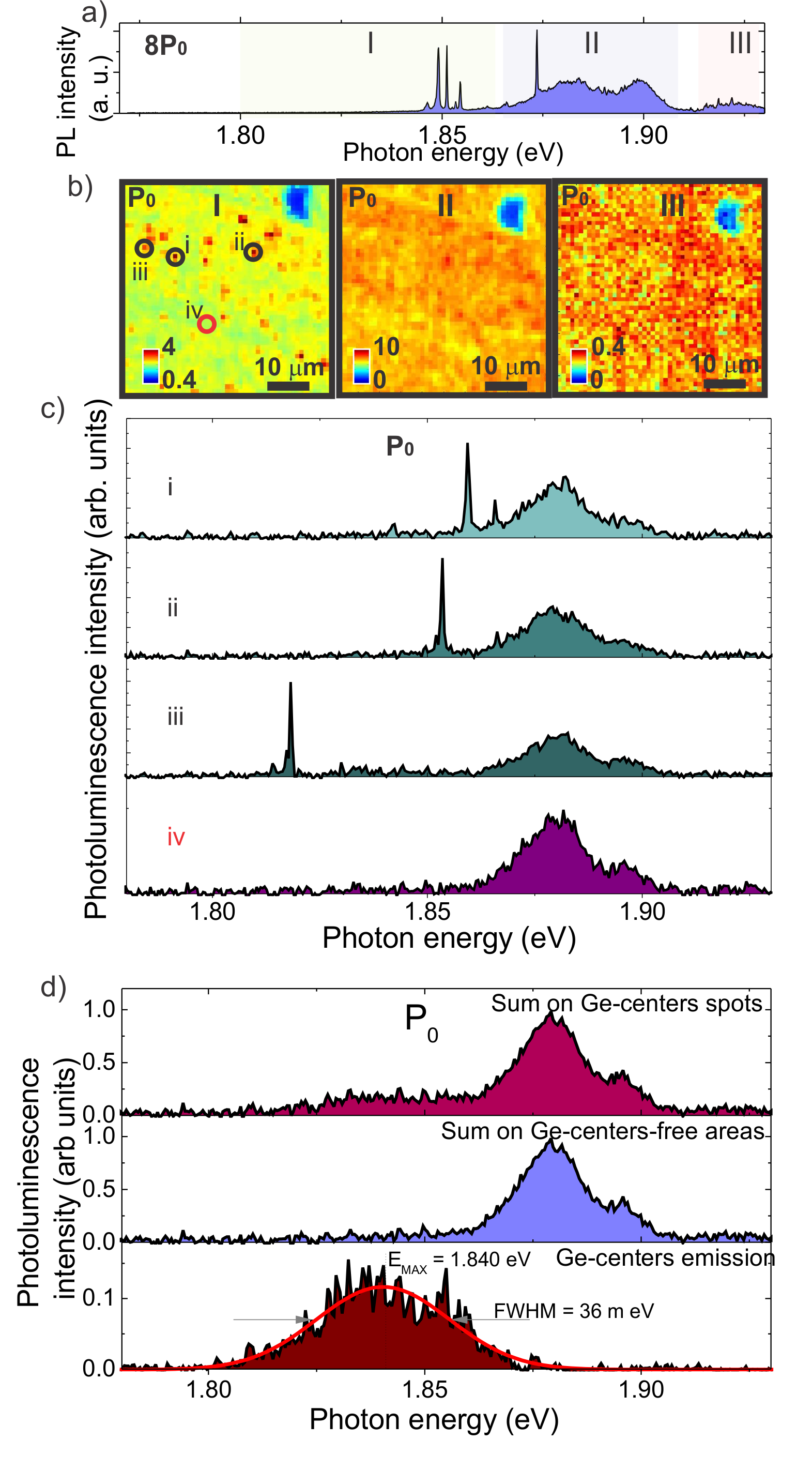}% Here is how to import EPS art
 \caption{  a) Typical PL spectrum from the sample Ge 580 $^\circ$C. The shaded areas I, II and III highlight the spectral intervals used to represent the spectral map shown in b). b)  
 $50 \times 50 \mu$m$^{2}$  spectral map  integrated on the three different spectral intervals I, II and  III highlighted in a).  The excitation power density was $P_{0} = 5 \times 10 ^{2}$ W/cm$^{2}$. 
 Similar maps but with a sharper spectral filtering are shown as an animated map in
 reference in the Supplementary Material \onlinecite{VideoMap1,VideoMap2}. In the left panel I four spots (i, ii, iii, and iv) are highlighted and the corresponding extracted spectra are shown in c).  d) Top panel: integrated and normalised emission from  Ge-centers. Central panel: same as top panel but for Ge-centers-free areas (i.e. ``bulk" Al$_{0.3}$Ga$_{0.7}$As). Bottom panel: Isolated Ge-centers emission obtained as intensity difference between the two spectra in top and middle panels. The continuous red line is a Gaussian fit to the data.\label{fig:Fig3}} 
 %b) Individual spectra  extracted from the
 %spatial map shown in b). The top three panels display the emission from
 %localised bright spots (left panel in b))  while the bottom one is from a
 %point without localised emission. \label{fig:Fig3}}
 \end{center}
 \end{figure}

We now focus the attention on  the case of the sample grown on a germanium substrate
by performing extended 2D PL maps at low excitation power. 
%Let us start by discussing the low power regime (laser intensity
%$\mathcal{P}_{0} = 40$ $\mu$W corresponding to an excitation power density of
%$P_{0} = 5 \times 10 ^{2}$  $W / $cm$^{2}$).
%While rastering the surface in  1 $\mu$m steps, we record a spectrum for
%each step integrating for 0.25 s, thus building  2D spectral-maps as large
%as $50 \times 50 \mu$m$^{2}$. For this kind of measurements, due to the
%large number of spectra, we limit the amount of data by  binning the energy
%scale over three pixels leading to a slightly reduced spectral resolution
%(about 300 $\mu$eV in full width at half maximum, which in turn, allows for
%a faster integration time with an improvement of the signal-to-noise ratio,
%together with a reduction of the amount of collected information; note that
%we register 2500 spectra for each map). Due to the limited spectral
%resolution used for mapping large area, we cannot access the real line
%broadening of the sharp lines from this analysis (see Section\ref{sec:LINE
%BROADENING}).
In   Fig. \ref{fig:Fig3} a) we highlight three spectral intervals: I  for Ge-impurities emission, II for the DA-Ge and DA-C emission and zone III for the BEx emission. 
In Fig.\ref{fig:Fig3} b) are displayed the results of a surface scan
spectrally integrated over the three  intervals highlighted in
Fig.\ref{fig:Fig3} a). We observe a strong spatial localization of the
emitted intensity from the spectral interval I (Fig.\ref{fig:Fig3} b) left
panel) in bright spots which lateral extension in space reflects our
instrumental resolution. For the zones II and III instead
(Fig.\ref{fig:Fig3} b) central and right  panels respectively), a rather
uniform intensity distribution is observed (see also the maps reported in the supplementary materials Supplementary Material \onlinecite{VideoMap1,VideoMap2}). 
The dark area in the top right part of the images is attributed to an extended defect of the crystal or to some impurity on the sample surface.
 
%Fig.\ref{fig:Fig3} b)  displays photoluminescence spectra relative to
%spatial coordinates of three bright spots  (first three spectra from top)
%and from a low intensity emission position where no localised emission is
%present (bottom spectrum). In the first class of spectra we always observe
%sharp peaks  below 1.87 eV  plus the broad double peak (in the spectral
%interval  II) attributed to the DA-Ge and -C emission.  In this low
%excitation regime  the sharp peaks are usually characterised by singlets or
%doublets of lines (with a few meV separation in energy) and are attributed
%to the Ge-impurities emission.\cite{Sarti2013}
%Similar maps but with a sharper spectral filtering are shown as animated
%maps in references \onlinecite{VideoMap1,VideoMap2} (in this case the 
%power was raised to  $8 P_{0}$).

The micro-PL mapping allows for a precise counting of the
emitting Ge-impurities within the investigated sample surface (and sample
thickness). Integrating over two maps of  50 $\times$ 50 $\mu$m$^{2}$
we find a surface density of spatially localised and sharp peaks of  about
$\sim 2.5 \times 10 ^{-2}\mu$m$^{-2}$. By taking into account the thickness
of the Al$_{0.3}$Ga$_{0.7}$As layer, we evaluate a concentration of $\rho
_{PL} \simeq 10^{11} $cm$^{-3}$
which is by far smaller than the concentration of Ge ions measured in SIMS
$\rho _{Ge} \sim  10^{17}$ cm$^{-3}$ (see Fig. \ref{fig:Fig1} b)).
 
The low density of quantum emitters and their PL-spectra with multiple lines  
is an indication of the complex nature of these extrinsic centers.
As for similar systems of extrinsic centers able to accommodate
multiexcitonic
states,\cite{Francoeur2004,Ikezawa2012,Zhang2013,Ethier2014,Jo2013}   we
believe that a likely attribution of the defects responsible for the
isolated emission below 1.87 eV can be the presence of binary systems or
more generally, complex defects. As an example, let us explicitly refer to
the attribution of N-dyads in AlAs as responsible of biexciton emission
\cite{Jo2013} and tentatively assume that our extrinsic centers were related
to pairs of Ge impurities. With this hypothesis and assuming a stochastic
model\cite{Thomas1966,Jo2013} it is possible to compare the measured density
of emitting centers ($\rho_{PL}$) with that of pairs of Ge ions
($\rho_{Ge-Ge} = \rho_{Ge}^{2} V$ where $V=d_{Ge-Ge}^{3}$ is the volume
occupied by a pair). We can roughly estimate the
typical distance between two ions $d_{Ge-Ge}$ in the $\sim$0.1 nm range
which well compares with the lattice constant of the host
semiconductor matrix. This  estimate strongly supports the idea
of extrinsic centers related to Ge-pair, even if different hypothesis cannot
be ruled out at the moment. Then in the following we will refer to them as
Ge-centers emission.

%The differences in the measured values of $\rho  ^{SIMS}$ and $\rho ^{PL}$
%may be due to local fluctuations affecting  either SIMS and
%photoluminescence measurements, but it may also reflect a non-uniform
%distribution of the Ge-doping (in the merit of this see Fig.\ref{fig:Fig6}
%and relative discussion) leading to a larger probability to find two ions at
%short distances. We mention that an overestimation of the number of  the
%emitting centers may also be originated from a non correct attribution of
%their emission due to the multi-excitonic features or by including some
%isolated, low emission energy,   DE-C and DE-Ge  centers. These latter
%contributions  can eventually lead to sharp features in the spectra similar
%to those previously shown\cite{Minari2012} with the main difference of the
%absence of multi-excitonic features which instead characterises
%impurity-pairs emission.\cite{Marcet2010a,Marcet2010,Jo2013}
 
Finally, we remark that this hyper-spectral imaging technique\cite{Bao2012}
allows the detection of complex dopant impurities\cite{Castelletto2014}  with an
extremely high sensitivity (note that the most accurate SIMS measurements
can reach $\sim 10 ^{13}$cm$^{-3}$). Differently from conventional
near-field microscopy techniques for doping detection,\cite{Castell2003} we
can gather additional information about the energy of each defect within the
energy band-gap of the host material and thus, \textit{a posteriori}, we can
re-build the full spectrum of the Ge-centers (see Fig. \ref{fig:Fig3}d)).

%   \begin{figure}[t]
%   \begin{center}\leavevmode
%   \includegraphics[width=1\linewidth]{Fig4.png}% Here is how to import EPSart
%   \caption{a) Top panel: integrated and normalised emission from  Ge-centers
%   at low excitation power density ($P_{0} = 5 \times 10 ^{2}$
%   W/cm$^{2}$). Central panel: same as top panel but for Ge-centers-free areas (i.e.
%   ``bulk" Al$_{0.3}$Ga$_{0.7}$As). Bottom panel: Isolated Ge-centers emission
%   obtained as intensity difference between the two spectra in top and middle
%   panels. The continuous red line is a Gaussian fit to the data.  b) same as
%   a) at a larger excitation power density $8P_{0}$.   \label{fig:Fig4}}
%   \end{center}
%   \end{figure}

%\section{$Ge$-CENTERS ENSEMBLE EMISSION}\label{sec:ENSEMBLE}

 %c) Top, central and bottom
% panels respectively show: Spectral shift of Ge-centers emission (relative to
% the low power $P_{0}$ case, $E_{0}=$1.8404 $eV$), line broadening and
% intensity  with increasing excitation power. The data are extracted from
% Gaussian  fits.
 
Combining the spectral information with the mapping over large areas of the
sample, a precise characterization of the ensemble emission can be extracted
despite the very low density of the Ge-centers. As a matter of fact standard
macro-PL measurements do not allow to extract the spectral emission band of
these Ge centers.\cite{Minari2012} The basic idea is to sum up only the micro-PL spectra arising from points containing the Ge-centers emission. This means
that within the two  50 $\times$ 50 $\mu$m$^{2}$ spectral maps, containing 2500
different points each, we operate a careful selection of 150
micro-PL spectra  corresponding to individual Ge-centers
emission. In this way we can build \textit{a
posteriori} a \textit{pseudo-macro-PL} spectrum by summing
them and normalising the total spectrum to its maximum intensity. 

The result of this operation is shown in the top panel (\textit{Sum on  Ge-centers
spots}) of Fig.\ref{fig:Fig3} d). The same procedure is
applied to micro-PL spectra where no localised and sharp
emission is present (\textit{Sum on  Ge-centers-free areas}), thus recovering the
``bulk"-like emission of  Al$_{0.3}$Ga$_{0.7}$As. This is shown in the central panels of Fig.\ref{fig:Fig3} d). 
Clearly the emission of the sharp lines is perfectly rejected in these
spectra where only DA-Ge, DA-C and Bex are visible.

%In the bottom panels of Fig.\ref{fig:Fig4} a) and b) (\textit{Ge defect
%emission}) it is shown the difference between the
%\textit{pseudo-macro-photoluminescence} spectra shown in the corresponding
%two top panels allowing for a clear separation of the Ge-centers emission
%spectrum.  The data from the two different maps are very similar and show a
%Gaussian-envelope with central emission at $\sim$1.840 eV with a FWHM
%of about $\sim$35 meV. It is worth stressing that our pseudo-macro PL
%spectra approach is able to retrieve important information on the spectral
%distribution of Ge-centers emitters, despite their very sparse density.

The large broadening of the PL band of the Ge-centers, may reflect either
random distribution of the ion-to-ion distance $d_{Ge-Ge}$ of the Ge-centers
in the host material or the spectral diffusion associated with stochastic
variations of the electric field\cite{Neuhauser2000,Berthelot2006,Coolen2007,Abbarchi2008,Mano2009,
Sallen2010,Abbarchi2011,Nguyen2012,Nguyen2013,Matthiesen2014}  and/or alloy around the Ge-centers. The large broadening is also in
stark contrast with what found in other systems like carbon\cite{Cloutier2005,Berhanuddin2012}  and copper\cite{Weber1982,Sumikura2014}  impurities in
silicon or nitrogen-pairs in AlAs or GaAs,\citep{Muller2006,Jo2013,Marcet2010}  featuring a
well defined emission energy reflecting a limited number of optically active configurations for the impurities 
while it looks similar to what was  found in Te-dyads in ZnSe,\cite{Muller2006,Marcet2010,Marcet2010a}  Mn
in ZnS,\cite{Sapra2005} and carbon-antisite pairs in SiC\cite{Castelletto2014} where a relevant spread in the emission energy is present. At the same time, this large variation of excitonic recombination can be an
advantage if a tuning or a selection of peculiar emission properties is
needed. Note also that the emission of Ge-centers is related to the band
gap of the host alloy, and then, in principle, a control of the emission
energy could be obtained by tuning the Al content or preliminary using
different III-V alloys. This is important for controlling the emission of
these quantum sources  to specific targets\cite{Akopian2011,Siyushev2014} or
for coupling them to other optical devices (such as optical fibers or
resonators). 

\section{MULTI-EXCITON EMISSION}\label{sec:HIGH POWER}

   \begin{figure}[t]
   \begin{center}\leavevmode
   \includegraphics[width=.9\linewidth]{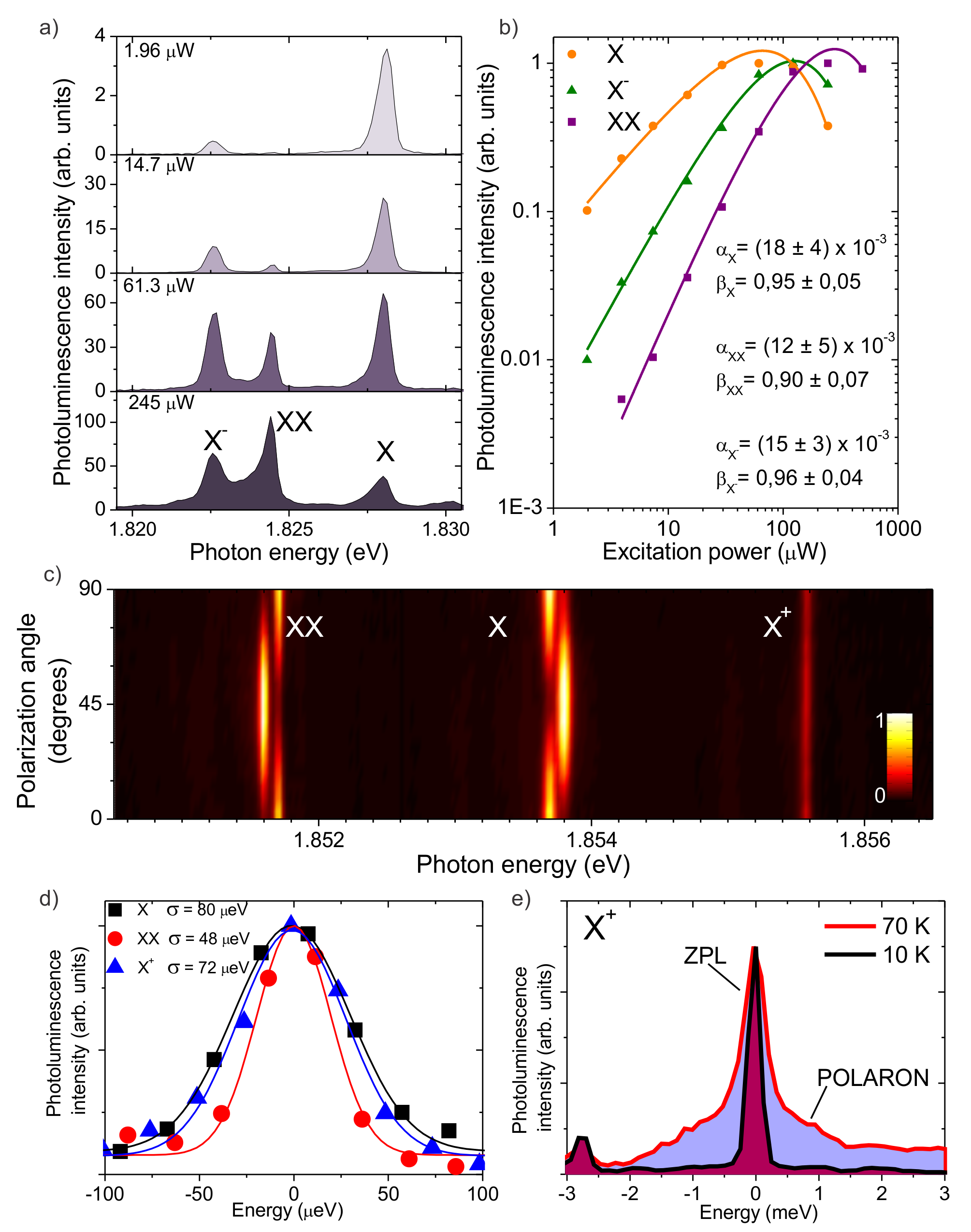}% Here is how to import EPS art
   \caption{ a) Series of PL spectra at different excitation 
   power of an individual Ge-impurity exhibiting X$^{-}$, XX and X recombination.
   b) Summary of the PL intensity as a function of the
   excitation power $\mathcal{P}$ of the data reported in a). Symbols are the experimental data while 
   the lines represent Poissonian fit. c) Polarization map of 
   an individual Ge-impurity exhibiting  XX, X and X$^{+}$ recombination.
   d) X, XX and X$^{+}$ normalized and energy- shifted emission at 10 K. The continuous lines are Gaussian fit to the data.  e) Normalised spectra of X$^{+}$ emission at 10 K and 70 K. The zero-phonon-line (ZPL) and the polaron emission are highlighted. \label{fig:Fig4}}
   %the spectral components of the s-shell (starting from the low energy side:
   %X$^{-}$, XX, X$^{+}$ and X. The PL from X$^{-}$ and X are multiplied by a
   %factor of 5 for a better visibility. \label{fig:Fig5}}
  \end{center}
   \end{figure}

As found for other complex defects\cite{Marcet2010,Marcet2010a} the
Ge-centers in study support charged excitons and multi-exciton features.
We note that, considering a statistical analysis of $\sim$150 Ge-centers, practically all of them (more
than 90$\%$) show these features and we thus conclude that the
Ge-centers in study support all the energy states typical of s-shell
recombination: up to two bound electrons and two holes (respectively in
conduction and valence band).  A preliminary assignment of
neutral exciton, biexciton and charged exciton was previously done by power
dependence and fine structure splitting measurements.\cite{Minari2012,Sarti2013}  
Here we complete the full picture of the electronic states of the s-shell 
addressing both the positive (X$^{+}$) and negative (X$^{-}$) charged exciton 
levels. At the same time, we evaluate the capture volume and the carrier 
localization through the analysis of the saturation power.\citep{Abbarchi2009}

In Fig.\ref{fig:Fig4} a) we show typical PL spectra of an individual Ge-center 
at different power density. The emission clearly shows
additional spectral components ascribed to biexcitons and charged exciton 
complexes.\cite{Sarti2013} The corresponding evolution of the PL intensities 
with power  is displayed in Fig.\ref{fig:Fig4} b). Here, according to a Poissonian 
model for the level occupation probability\cite{Abbarchi2009}  the three main PL lines 
follow slightly different filling dynamics (the fits shown in Fig.\ref{fig:Fig4} b)
correspond to $I_{PL} \propto (\alpha \mathcal{P} ^{\beta}) ^{m}  \exp (-(\alpha \mathcal{P} ^{\beta}) ^{m} ) $ where $m = 1, 1.5, 2$ corresponds respectively to one exciton (X), one exciton plus a spectator charge  (X$^{\pm}$) and two excitons (XX); $\mathcal{P}$ is the excitation power; $\alpha$ and $\beta$ are fitting parameters describing the capture mechanism and their corresponding values extracted from the fit are directly reported on Fig.\ref{fig:Fig4} b)). The dependence 
on the excitation power,\cite{Abbarchi2009}  the line broadening,\cite{Abbarchi2008} and the relative spectral position\cite{Abbarchi2010} leads us to the attribution of the lines to  X$^{-}$, XX,  and X. The X line reported in Fig. \ref{fig:Fig4} b) shows a
saturation power of about $\mathcal{P}_{sat} \simeq$ 0.6 mW  
while slightly different values (not shown) were observed for
other Ge-centers (from $\sim$0.05 up to $\sim$1
mW). This feature possibly reflects a different capture efficiency due to
the presence of extrinsic effects related to disorder in the electrostatic
environment of the emitters.\cite{Abbarchi2009,Sallen2010,Abbarchi2012,Nguyen2012,Nguyen2013,Matthiesen2014} 

The saturation power of X (i.e. the value $\mathcal{P} _{sat}$ for the maximum occupation probability) allows to estimate the capture length of the Ge-centers.\cite{Abbarchi2009} From the typical values of $\mathcal{P} _{sat}$ for X and  taking into account the experimental conditions, we estimate a capture length  ranging from $\sim 30$ nm up to  $\sim 10$ nm, similar or smaller of that found for  GaAs/Al$_{0.3}$Ga$_{0.7}$As epitaxial quantum
dots\cite{Abbarchi2009,Cavigli2012,Accanto2013} investigated with the same
experimental apparatus and with similar excitation conditions.

Clearcut attribution of the exciton-biexciton cascade and charged exciton comes from
fine structure splitting measurements: an example of a different individual Ge-center
showing  X$^{+}$, XX,  and X is reported in Fig.\ref{fig:Fig4} c). The X and XX lines shifts symmetrically when changing the detected polarization angle, while the  X$^{+}$ does not shift.\cite{Rodt2007,Abbarchi2008,Abbarchi2009,Liu2014}

%Generally speaking the assignment of positive and negative charged
%exciton X$^{+}$ and X$^{-}$ is tentatively done by considering the expected
%larger binding energy for the complex X$^{-}$.

%This attribution also agrees with the line broadening measurements (see reference \onlinecite{Abbarchi2008a}): while  the linewidth of X$^{-}$ is similar to that of X, 
%X$^{+}$ is usually sharper. 

Let us now consider the line broadening of the different recombination lines as 
an additional key for attributing the excitonic complexes (see Fig.\ref{fig:Fig4} d). 
From a statistical point of view we observe a wide spread of values of the
linewidth ranging from few tens up to few hundreds of  $\mu$eV, still the
order from the sharper to the broader lines among the different excitonic
complexes does not vary: $\sigma _{X^{-}} \sim \sigma _{X} > \sigma _{X^{+}}
> \sigma _{XX}$ (for the sake of thoroughness we note that X$^{-}$ featuring larger or smaller broadening with respect to the corresponding X can be found).  
The gaussian broadening of the low temperature
PL emission of individual excitonic transitions is commonly
ascribed to spectral diffusion and
\cite{Neuhauser2000,Berthelot2006,Coolen2007,Abbarchi2008,Mano2009,
Sallen2010,Abbarchi2011,Nguyen2012,Nguyen2013,Matthiesen2014} a
hierarchical broadening (showing $\sigma _{X^{-}} \sim \sigma _{X} > \sigma
_{X^{+}} \sim \sigma _{XX}$)  has been already found in epitaxial quantum
dots\cite{Abbarchi2008,Mano2009,Adachi2007,Kumano2008} and
impurities.\cite{Corfdir2014}
This  behaviour can be  explained in terms of a \textit{selective} quantum
confined Stark effect induced by the presence of charged defects in the
semiconductor matrix.\cite{Abbarchi2008,Mano2009} The relative position of
the charged trap results in a different Stark shift amplitude (and
eventually sign) for each  excitonic complex.  On one side, the analogy with
QDs is expected for any strongly confining potential able to localize the
electrons and holes at the nm scale, as in the case of the Ge-centers in
study. On the other side, we did use  this analogy with QDs in order to
attribute the emission lines to different excitonic complexes. In this
respect the hierarchical broadening is a confirmation of the soundness of
our attribution.

In view of possible integration in optoelectronic devices it is also needed
to increase the operation temperature of the quantum sources as much as
possible. Indeed we were able to follow the emission of the Ge-centers up to 70 K. 
Following the temperature-induced red-shift of any excitonic recombination
we observe an initial linear decrease in energy and then a steeper quadratic
decrease (as reasonably expected for electronic transitions in
semiconducting materials according to the Varshni empirical law of AlGaAs
alloy).\cite{varshni1967,Sanguinetti2006} 
While at low temperature only a sharp line is visible, with a spectral
broadening limited by our low resolution (in these data) at higher
temperature we assist to a slight  broadening of the central sharp line
together with the onset of a broad
pedestal.\cite{Takagahara1999,Krummheuer2002,Favero2003,Muljarov2004,
Berthelot2006,Abbarchi2008a} These features can be interpreted in terms of
electron-acoustic phonon interaction: in analogy with epitaxial quantum
dots, we assign the sharp component of the single Ge-impurities emission to
the exciton zero phonon line (ZPL in Fig. \ref{fig:Fig4}e) while the
sidebands  to a superposition of acoustic phonon replicas. The still sharp
emission at T = 70 K and the limited  thermal quenching of the PL (quite
similar to what observed in GaAs QDs) lead us to state the possibility to
use the Ge-centers as quantum emitters at liquid nitrogen temperature.

\section{BINDING ENERGY OF XX, X$^{+}$ AND  X$^{-}$}\label{sec:BINDING
ENERGY}

We now address the issue of the Coulombic interactions between electrons and
holes trapped by Ge-impurities which are responsible for the lifted
degeneracy of the neutral exciton state as well as for the energy
re-normalization of the electronic states within individual impurity centers. 

Due to asymmetries in the confinement potential and local strain accumulation 
the two spin states of the bright neutral exciton are split by the so called fine 
structure splitting (FSS) associated to the exchange interaction;\cite{Rodt2007,Abbarchi2008c,Mano2010,Kuroda2013,Liu2014,Ha2014,Jo2013} for the Ge-centers 
in study, it has been shown that the FSS is in the $\sim$100 $\mu$eV
range (see Fig.\ref{fig:Fig4} c)).\cite{Sarti2013} 

The so called exciton binding energy is simply defined as the 
energy shift with respect to the corresponding neutral
exciton line ($BE_{XX} = E_{X}-E_{XX}$). Similarly, one can define the
charged exciton binding energy as $BE_{X^{\pm}} =
E_{X}-E_{X^{\pm}}$). The binding energy
arises form the complicate interplay of direct Coulombic interactions,
exchange and correlation typical of each excitonic
species.\cite{Franceschetti1999,Tsuchiya2000,Regelman2001,Filinov2004,
Moody2013,Abbarchi2010,Marcet2010,Rodt2005,Schliwa2009,Moody2013,Liu2014}

  \begin{figure}[t]
  \begin{center}\leavevmode
  \includegraphics[width=.7\linewidth]{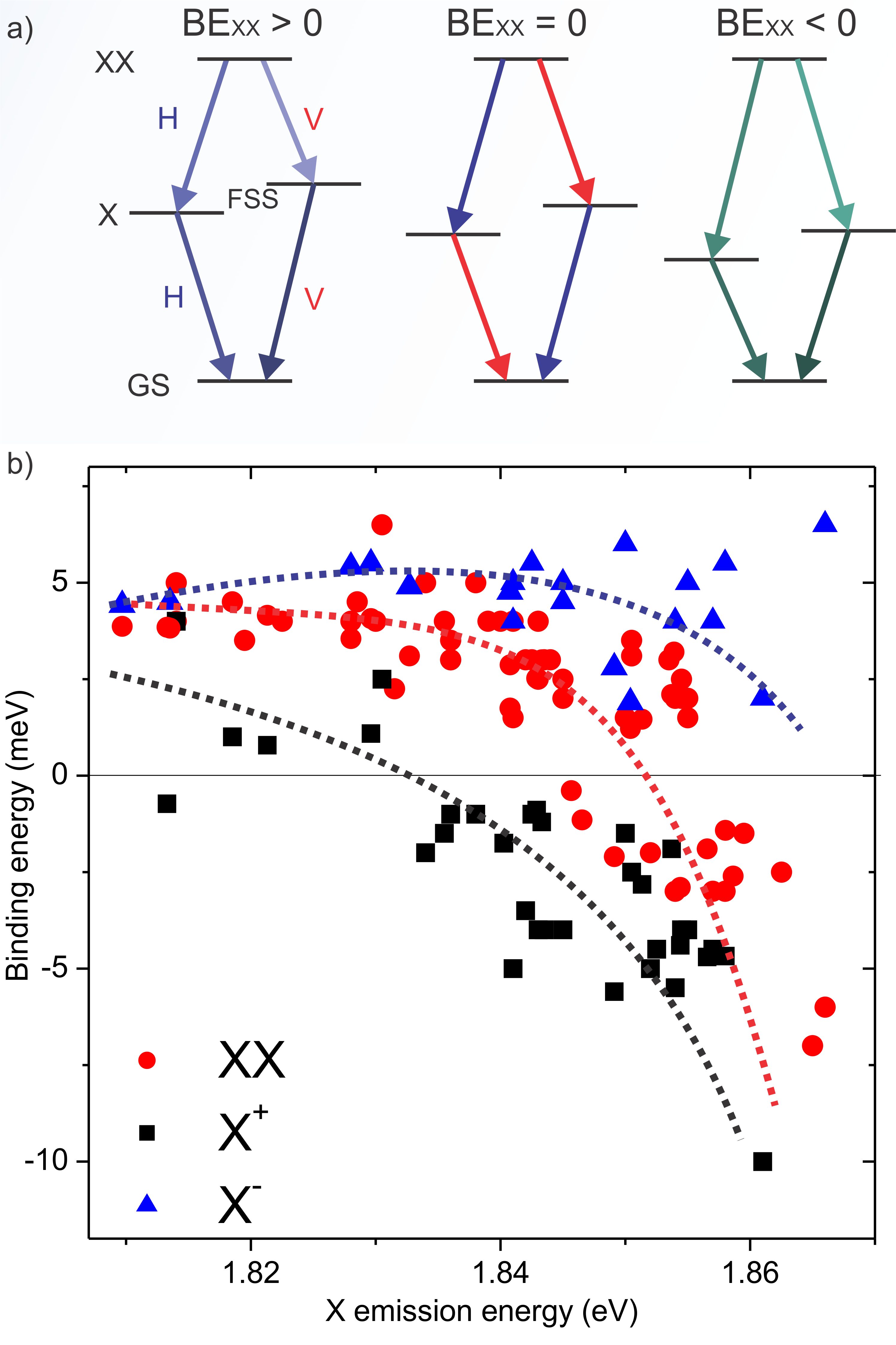}% Here is how to import EPS art
  \caption{ a) Scheme of X and XX energy levels for three different binding
  energy of XX; from left to right: positive, zero, and negative. V and H highlight  the linear polarization of the emitted photons. b)  Binding energy of X$^{-}$, X$^{+}$ and XX. Dotted lines are guides to the eyes.
  \label{fig:Fig5}}
  \end{center}
  \end{figure}

Generally speaking three scenarios can occur, the binding energy is positive,
zero or negative.
In the case of the biexciton complex these cases are represented in Fig.
\ref{fig:Fig5} a) also including the presence of FSS (not in scale). In the
left panel of Fig. \ref{fig:Fig5} a)  is displayed  the situation of a XX
having positive binding energy $BE_{XX} > 0$.  In this condition (if FSS
$\neq $0) the emitted photons in the biexciton cascade are discernible in
energy and polarization. If instead the FSS vanishes, the XX and X photons are
polarization-entangled.\cite{Chen2000,Li2003,Stevenson2006,Akopian2006,Kuroda2013,Jayakumar2013,Trotta2012a,Trotta2012,Trotta2014} In the central panel of  Fig. \ref{fig:Fig5} a)  is shown a XX
with a zero binding energy $BE_{XX} = 0$ where the emission energy of
H-(V)polarised XX   photons matches that of the V-(H) polarised  X photon.
The two photons are not distinguishable in energy and thus polarization
entanglement can be recovered with a time-reordering
scheme.\cite{Troiani2008,Avron2008} Finally, in the right panel, it is shown
the case of negative binding energy ($BE_{XX} < 0$); here the same considerations
made for the left panel apply.
Therefore, in view of realising a quantum source of entangled photons pairs, the
condition can be $FSS = 0$ or $BE_{XX} = 0$.

Let us know consider the experimental data for the binding energies of
excitonic complex localised at the Ge-centers. As most of these transitions
were mainly attributed on the base of the power dependence it is usually
easy to have a clear attribution of  X and  XX while the X$^{+}$ and X$^{-}$
attributions are sometimes less certain. Disregarding the ambiguous cases
leads to a lower statistic for X$^{+}$ and X$^{-}$. Summary of the data with
excitonic recombination ranging from $\sim$1.81 eV to $\sim$1.86 eV is given in Fig.
\ref{fig:Fig5} a). Pronounced features emerging from this analysis are summarized as follows:
\\
-the binding energy of X$^{+}$ shows a decreasing trend when increasing X
emission energy, from $\sim$+1 meV to $\sim$-5 meV, with a transition from
binding  to  anti-binding. \\
-The XX state shows a binding energy  of about $\sim$+4 meV at low energy and
abruptly evolves toward negative values above 1.855 eV reaching the negative
value of about $\sim$-6 meV with a steep trend.  \\
-The binding energy  of X$^{-}$ is always positive, thus X$^{-}$  forms a
bound state. Its evolution with the 3D confinement is quite constant even if
the number of points is quite small due to the problematic attribution of
this line.
 
Comparing these results with calculations and experimental data for  quantum
dots or monolayer fluctuation of quantum wells
\cite{Tsuchiya2000,Regelman2001,Filinov2004,Moody2013,Abbarchi2010,
Abbarchi2010a} we can see a remarkable agreement with the behaviour predicted and
measured for a truly 3D confinement\cite{Tsuchiya2000,Abbarchi2010} (i.e. a
small quantum dot without a wetting layer).  XX and X$^{+}$ show a steep
reduction in the binding energy and eventually a change of its sign. This is
interpreted  as a consequence of the dominant exciton mean-field interaction
present in a spherical confining well, reflecting a strong carrier
localization in all the spatial directions. Nevertheless, as a consequence of the larger
effective mass of holes with respect to electrons, the same mean field would
serve as a repulsive force for a hole and an attractive one for electrons 
thus leading to an opposite trend and sign for X$^{+}$ and X$^{-}$ binding energies.

These findings are in contrast with  common epitaxial quantum dots with a weaker lateral
confinement originated by a 2D wetting layer: in these cases a linear dependence of
the XX and X$^{+}$ binding energy as a function of the X emission is usually
found together with  a dominantly binding-nature of the XX
state.\cite{Abbarchi2010,Abbarchi2010a,Rodt2005,Schliwa2009,Moody2013}
 
We conclude that a strong carrier localization is present for high energy
Ge-centers, while the positive values of the  binding energy found at lower
X emission energy suggests a recovered dominant effect of the correlation
term. Most importantly, a XX state emitting at energies close to that of the
corresponding X state can be achieved, thus allowing for an easier
implementation of a \textit{time reordering} scheme for entangled
photons\cite{Troiani2008,Avron2008}  eventually with the help of a (small)
external  field for the fine tuning.\cite{Trotta2012a,Trotta2012,Trotta2014}

%\textit{[PROBLEM:] The measured data four our Ge-centers show an opposite
%trend with respect to the measurements shown in \onlinecite{Marcet2010} for
%tellurium  in ZnSe isoelectronic centers. [IN THIS CASE THE "emission energy
%increases with the inter-
%atomic separation of the centers"!!! I DO NOT UNDERSTAND] }

\section{CONCLUSIONS}\label{sec:CONCLUSIONS}

In conclusion we have shown that the single photon emitters recently 
reported\cite{Minari2012,Sarti2013} are related to Ge-centers and 
come from unintentional contamination of the III-V layers grown on germanium substrates. 
These centers provide a valuable
alternative to epitaxial quantum dots for the implementation of different
carrier states such as X, XX, X$^{+}$ and  X$^{-}$ and, in analogy with 
similar impurities, could be in principle implemented through ionic implantation.\cite{Cloutier2005,Berhanuddin2012,Weber1982,Sumikura2014,Castelletto2014,Lesik2013} 
Due to the strong carrier confinement the Ge-centers feature similar properties to the
epitaxial quantum dot counterpart, present a line broadening dominated by a
quantum-confined Stark effect at low temperature and a phonon-coupling at
higher temperatures. The optical properties of the excitonic recombination
from Ge-centers are quite good: linewidth as sharp as 40 $\mu$eV can be
found at high excitation power densities and low temperature, the emission
is still bright at 70 K, confirming the good thermal stability of this class
of emitters.  The electronic states populating the s-shell well agree with
what found for a fully 3D confining potential in quantum dots and thus
enable the implementation of quasi degenerate X and XX states. 
 
Our findings suggest the use of the PL emission of Ge
extrinsic centers in Al$_{0.3}$Ga$_{0.7}$As  as a versatile platform for
obtaining single photons  on a large spectral range and entangled photons
based on the \textit{time reordering scheme},\cite{Troiani2008,Avron2008}
for spin-photon turn-stile devices and for slow-light when coupled with
atomic vapours.\cite{Akopian2011,Siyushev2014}  We stress that, differently
from most III-V compounds where Stranski-Krastanov quantum dots are obtained,
the emission of these Ge-centers well matches the highest detection
efficiency of Si-based single photon detectors ensuring a high
fidelity in the detection and not only in the preparation of the quantum
state of the system.\cite{Eisaman2011} Last but not least,  these results
have been obtained on silicon and germanium substrates opening up new
avenues for the exploitation of quantum emitters within a device-friendly
platform.

\begin{acknowledgments}

This work has been carried out thanks to the support of the European project 
\textit{LASERLAB-EUROPE}  (grant agreement no. 284464, EC's Seventh Framework).

\end{acknowledgments}

%\begin{thebibliography}{99}
%
%\bibitem{KRK06}
%J. Kasprzak, M. Richard, S. Kundermann, A. Baas, P. Jeambrun, J. M. J. Keeling, F. M. Marchetti, 
%M. H. Szymanska, R. Andre, J. L. Staehli, V. Savona, P. B. Littlewood, B. Deveaud, and Le Si Dang, Nat. \textbf{443}, 409 (2006); D. Sanvitto, A. Amo, L. Viña, R. André, D. Solnyshkov, and G. Malpuech Phys. Rev. B \textbf{80}, 045301 (2009)
%
%\end{thebibliography}

\bibliography{Ge_dyads_MG}
\bibliographystyle{apsrev4-1}

%%%%%%%%%%%%%%%%%%%%%%%%%%%%%%%%%%%%%%%%%%%%%%%%%%%%%%%%%%%%%%%%%%%%%%%%%%%%%%%%%%%%%%%%%%%%%%%

%%%%%%%%%%%%%%%%%%%%%%%%%%%%%%%%%%%%%%%%%%%%%%%%%%%%%%%%%%%%%%%%%%%%%%%%%%%%%%%%%%%%%%%%%%%%%%%

%%%%%%%%%%%%%%%%%%%%%%%%%%%%%%%%%%%%%%%%%%%%%%%%%%%%%%%%%%%%%%%%%%%%%%%%%%%%%%%%%%%%%%%%%%%%%%%

%%%%%%%%%%%%%%%%%%%%%%%%%%%%%%%%%%%%%%%%%%%%%%%%%%%%%%%%%%%%%%%%%%%%%%%%%%%%%%%%%%%%%%%%%%%%%%%
\newpage

%
%\begin{figure*}[t]
%\begin{center}\leavevmode
%\includegraphics[width=1\linewidth]{Fig1.png}% Here is how to import EPS art
%\caption{(Color online).\label{fig:Fig1}}
%\end{center}
%\end{figure*}
%
%\newpage
%
%\begin{figure*}[t]
%\begin{center}\leavevmode
%\includegraphics[width=.8\linewidth]{Fig2.png}% Here is how to import EPS art
%\caption{(Color online).\label{fig:Fig2}}
%\end{center}
%\end{figure*}
%\newpage
%
%\begin{figure*}[t]
%\begin{center}\leavevmode
%\includegraphics[width=.7\linewidth]{Fig3.png}% Here is how to import EPS art
%\caption{(Color online).\label{fig:Fig3}}
%\end{center}
%\end{figure*}
%
%\newpage
%
%\begin{figure*}[t]
%\begin{center}\leavevmode
%\includegraphics[width=.7\linewidth]{Fig4.png}% Here is how to import EPS art
%\caption{(Color online).\label{fig:Fig4}}
%\end{center}
%\end{figure*}
%
%\newpage
%
%
%\begin{figure*}[t]
%\begin{center}\leavevmode
%\includegraphics[width=1\linewidth]{Fig5.png}% Here is how to import EPS art
%\caption{(Color online).\label{fig:Fig5}}
%\end{center}
%\end{figure*}
%
%
%\newpage
%
%
%\begin{figure*}[t]
%\begin{center}\leavevmode
%\includegraphics[width=1\linewidth]{Fig6.png}% Here is how to import EPS art
%\caption{(Color online).\label{fig:Fig6}}
%\end{center}
%\end{figure*}
%
%
%\newpage
%
%
%\begin{figure*}[t]
%\begin{center}\leavevmode
%\includegraphics[width=.7\linewidth]{Fig7.png}% Here is how to import EPS art
%\caption{(Color online).\label{fig:Fig7}}
%\end{center}
%\end{figure*}
%
%
%\newpage

%\listoffigures

\end{document}